\documentclass[10pt]{wlscirep}
\usepackage[utf8]{inputenc}
\usepackage[T1]{fontenc}
\usepackage{bm}
\usepackage{amsmath}
\usepackage{siunitx}
\usepackage{caption}
\usepackage{lineno}
\usepackage{url}
\usepackage{xcolor}
\usepackage{mathrsfs}

\usepackage[acronym]{glossaries}  
\newacronym{cmos}{CMOS}{Complementary Metal-Oxide-Semiconductor}
\newacronym{tia}{TIA}{Transimpedance Amplifier}
\newacronym{relu}{ReLU}{Rectified Linear Unit}
\newacronym{dtw}{DTW}{Dynamical Time Wrapping}
\newacronym{mre}{MRE}{Mean Relative Error}
\newacronym{ml}{ML}{Machine Learning}
\newacronym{ad}{A/D}{Analogue-Digital}
\newacronym{eds}{EDS}{Energy-dispersive X-ray Spectroscopy}
\newacronym{tem}{TEM}{Transmission Electron Microscope}
\newacronym{b1500}{B1500A}{B1500A Semiconductor Device Analyzer}
\newacronym{ode}{ODE}{Ordinary Differential Equation}
\newacronym{resnet}{ResNet}{Residual Network}
\newacronym{node}{NODE}{Neural Ordinary Differential Equation}
\newacronym{odenet}{ODE-Net}{Neural ODEs}
\newacronym{mts}{MTS}{Multivariate Time Series}
\newacronym{wl}{WLs}{Word Lines}
\newacronym{bl}{BLs}{Bit Lines}
\newacronym{sl}{SLs}{Source Lines}
\glsdisablehyper

\newcommand{\figref}[1]{Fig.~{\ref{#1}}}

\title{Continuous-Time Digital Twin with Analogue Memristive Neural Ordinary Differential Equation Solver}

\author[1,2,3,$^{\dagger}$]{Hegan Chen}
\author[1,2,3,$^{\dagger}$]{Jichang Yang}
\author[2,$^{\dagger}$,*]{Jia Chen}

\author[1,2,3]{Songqi Wang}
\author[1,2]{Shaocong Wang}
\author[1,2]{Dingchen Wang}
\author[1]{Xinyu Tian}
\author[1,2]{Yifei Yu}
\author[1,2]{Xi Chen}
\author[1]{Yinan Lin}
\author[1,2]{Yangu He}
\author[1,2]{Xiaoshan Wu}
\author[1,3,4,5]{Yi Li}
\author[1,2,3]{Xinyuan Zhang}
\author[1,2]{Ning Lin}
\author[1]{Meng Xu}

\author[6]{Yi Li}
\author[7]{Xumeng Zhang}
\author[1,2,3,*]{Zhongrui Wang}
\author[1,3,*]{Han Wang}
\author[4,5,*]{Dashan Shang}

\author[7]{Qi Liu}
\author[2,8]{Kwang-Ting Cheng}
\author[4,7]{Ming Liu}

\affil[1]{Department of Electrical and Electronic Engineering, the University of Hong Kong, Hong Kong, China}
\affil[2]{ACCESS – AI Chip Center for Emerging Smart Systems, InnoHK Centers, Hong Kong Science Park, Hong Kong, China}
\affil[3]{Institute of the Mind, the University of Hong Kong, Hong Kong, China}
\affil[4]{Key Laboratory of Microelectronic Devices \& Integrated Technology, Institute of Microelectronics, Chinese Academy of Sciences, Beijing 100029, China}
\affil[5]{University of Chinese Academy of Sciences, Beijing 100049, China}
\affil[6]{School of Integrated Circuits, Hubei Key Laboratory for Advanced Memories, Huazhong University of Science and Technology, Wuhan 430074, China.}
\affil[7]{State Key Laboratory of Integrated Chips and Systems, Frontier Institute of Chip and System, Fudan University, Shanghai 200433, China}
\affil[8]{Department of Electronic and Computer Engineering, the Hong Kong University of Science and Technology, Hong Kong, China}

\affil[$^{\dagger}$]{These authors contributed equally.}
\affil[*]{e-mail: zrwang@eee.hku.hk;jjiachen@ust.hk;hanwang@eee.hku.hk;shangdashan@ime.ac.cn}

\begin{abstract}
Digital twins, the cornerstone of Industry 4.0, replicate real-world entities through computer models, revolutionising fields such as manufacturing management and industrial automation.
Recent advances in machine learning provide data-driven methods for developing digital twins using discrete-time data and finite-depth models on digital computers. However, this approach fails to capture the underlying continuous dynamics and struggles with modelling complex system behaviour. Additionally, the architecture of digital computers, with separate storage and processing units, necessitates frequent data transfers and \gls{ad} conversion, thereby significantly increasing both time and energy costs.
Here, we introduce a memristive neural ordinary differential equation (ODE) solver for digital twins, which is capable of capturing continuous-time dynamics and facilitates the modelling of complex systems using an infinite-depth model. By integrating storage and computation within analogue memristor arrays, we circumvent the von Neumann bottleneck, thus enhancing both speed and energy efficiency.
We experimentally validate our approach by developing a digital twin of the HP memristor, which accurately extrapolates its nonlinear dynamics, achieving a 4.2-fold projected speedup and a 41.4-fold projected decrease in energy consumption compared to state-of-the-art digital hardware, while maintaining an acceptable error margin. Additionally, we demonstrate scalability through experimentally grounded simulations of Lorenz96 dynamics, exhibiting projected performance improvements of 12.6-fold in speed and 189.7-fold in energy efficiency relative to traditional digital approaches.
By harnessing the capabilities of fully analogue computing, our breakthrough accelerates the development of digital twins, offering an efficient and rapid solution to meet the demands of Industry 4.0.
\end{abstract}
\begin{document}
\flushbottom
\maketitle
\section*{Introduction}

Digital twins are computational models that dynamically evolve to deliver precise representations of the structure, behaviour, and context of specific physical assets, including components, systems, and processes. As the cyberspace counterpart of physical entities, digital twins are fundamental to the Industry 4.0. Recent advances, exemplified by the NVIDIA Omniverse\cite{O1,O2}, incorporate specialised hardware to streamline operations, reduce operational costs, and boost productivity\cite{O3,O4}, revolutionising fields such as manufacturing management and industrial automation\cite{DT1,DT2,DT3,DT4,DT5,DT6,DT7,DT8}.\par
Despite considerable advances, the development and deployment of larger and more sophisticated digital twins encounter significant challenges relating to data, model, and architecture\cite{DT_lar, DT_ML1, DT_ML2, DT_ML3}. 
Data-wise, the majority of digital twins employ discrete-time numerical methods to approximate time-continuous dynamics observed in the real world. While this approach is well-suited for digital computers, it inherently introduces truncation errors and information loss due to the sampling of continuous signals\cite{c1_1,c1_2,c1_3}. 
Model-wise, the ability of AI-powered digital twins to represent complex dynamics scales with the depth of machine learning models. However, efforts to enhance these capabilities by increasing model depth lead to significant growth in parameter population and training cost\cite{c2_1,c2_2}. 
Architecture-wise, traditional digital twins face significant challenges: frequent analogue-digital conversions of sensory data introduce substantial latency and elevate energy consumption\cite{hybrid}; the von Neumann bottleneck, arising from the physical separation of memory and processing units\cite{von1, von2, von3}, incurs significant time and energy overheads due to massive data shuttling; and the ongoing miniaturisation of \gls{cmos} devices is approaching the physical limits of Moore's Law, making CMOS scaling cost ineffective\cite{moore1, moore2}.\par
To address the aforementioned challenges, we propose an innovative solution: a continuous-time and in-memory neural ordinary differential equation (ODE) solver for infinite-depth digital twins. Our approach offers several advantages over conventional digital twins.
Data-wise, our digital twin handles signals from the physical world in a time-continuous manner. This approach effectively eliminates temporal information loss and truncation errors typically associated with models that rely on discrete time steps\cite{s1_1,s1_2,s1_3}.
Model-wise, to address the limitations of finite-depth networks\cite{s2_1,diffusion}, we employ neural ODEs to parameterize the derivative of the hidden state in continuous time\cite{neural_ode}. This framework offers an infinite-depth approximation to residual neural networks (ResNets), outperforming recurrent neural networks with the same parameter population and streamlining the training process\cite{s2_2,s2_3,s2_4,s2_5}.
Architecture-wise, our system leverages emerging memristors, known for their scalability and three-dimensional (3D) stackability\cite{s3_RRAM_1, s3_RRAM_2}, which potentially mitigate the slowdown of Moore's Law. Additionally, memristor arrays perform in-memory vector-matrix multiplications in analogue domain, leveraging Ohm's Law for multiplication and Kirchhoff's Current Law for summation. This offers a highly parallel and energy-efficient solution to address the von Neumann bottleneck. Moreover, this fully analogue system circumvents the additional energy and delay typically associated with analogue-digital conversions\cite{s3_im_1,s3_im_2,s3_im_3,s3_im_4}.\par
In this article, we validated our approach using \SI{180}{\nano\meter} integrated analogue memristor arrays for two different digital twins: the HP memristor with external inputs and Lorenz96 dynamics as an autonomously evolving system.
Compared to conventional digital twins on digital hardware, for HP memristor modelling, our digital twin achieves a remarkable 4.2-fold projected increase in speed and a substantial 41.4-fold projected enhancement in energy efficiency, while maintaining comparable absolute errors.
Moreover, our digital twin demonstrates exceptional scalability in simulated atmospheric pressure extrapolation based on the Lorenz96 dynamics\cite{lo1,lo2}, exhibiting a 12.6-fold projected improvement in speed and a 189.7-fold projected increase in energy efficiency. Notably, our system also shows superior robustness, demonstrating enhanced resistance to the inherent noise of analogue memristor arrays\cite{noise}.
Our time-continuous, infinite-depth, and in-memory analogue system lays the foundation for the next generation digital twins.

\section*{Digital twin using memristive neural ODE solver}
\begin{figure}[!t]
    \centering
    \includegraphics[width=0.9\linewidth]{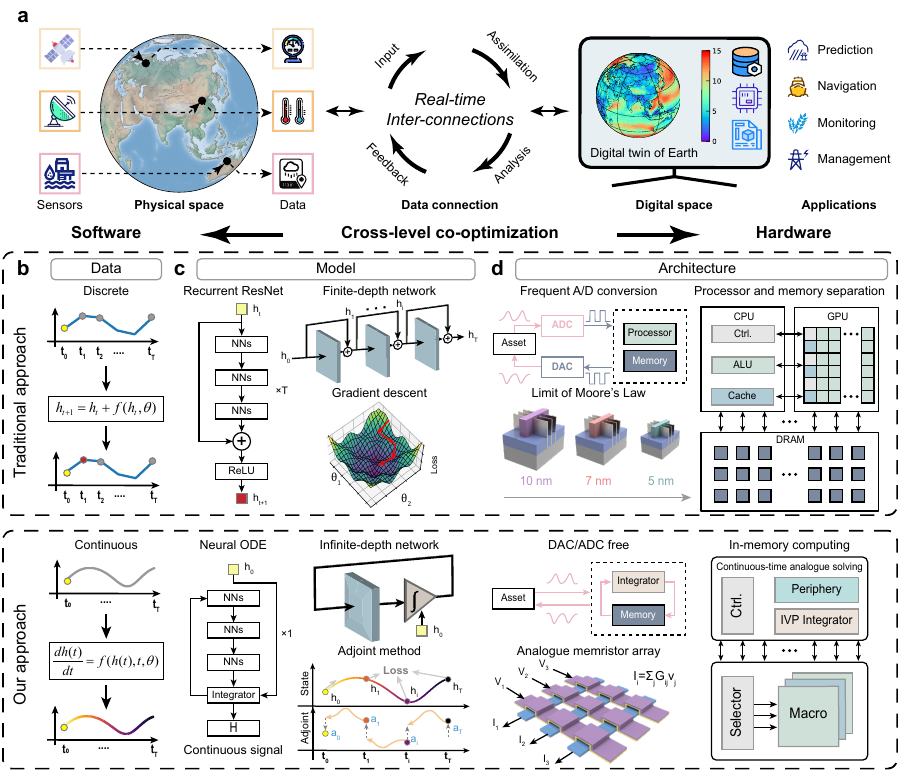}
    \caption{\textbf{A digital twin using a memristive neural ordinary differential equation (ODE) solver compared to a recurrent ResNet-based digital twin on conventional digital hardware in terms of data, model, and architecture}. 
    \textbf{a, Background,} The digital twin is a virtual counterpart of a physical asset. For example, environmental sensors capture data on atmospheric pressure, temperature, and precipitation, which is received by the digital twin. The digital twin models the dynamics of the physical system (e.g., atmospheric physics in this example).
    \textbf{b, Data,} (upper) The discretisation of continuous-time signals introduces truncation errors. (bottom) In contrast, our digital twin, using a memristive neural ODE solver, is intrinsically time-continuous.
    \textbf{c, Model,} (upper) Recurrent ResNet, by stacking neural network (NN) blocks, is a finite-depth network that parameterize a single discrete time transition using the ResNet. The representational capability scales with the depth, along with parameter population and training overhead. (bottom) In contrast, we employ a neural ODE that is equivalent to an infinite-depth neural network, featuring better performance and low training cost.
    \textbf{d, Architecture,} (upper) Conventional digital systems with frequent analogue-digital (A/D) conversion. The energy efficiency and parallelism are limited by the von Neumann bottleneck and the slowdown of Moore's Law. (bottom) In contrast, the in-memory neural ODE solver using the memristor array enables fully analogue computing without frequent data shuttling or the need for analogue-digital conversion.
    }
    \label{fig:co}
\end{figure}
In \figref{fig:co}, we illustrate the implementation of a digital twin using a memristive neural ODE solver and compare it with a recurrent ResNet on digital hardware in terms of data, model, and hardware architecture.\par
\figref{fig:co}a illustrates the interconnections between a physical asset and its digital twin, both of which are dynamic systems evolving within their respective state spaces. The digital twin acquires sensory data, such as those from satellites\cite{DT4}, and this data stream updates the state of the digital twin, ensuring an accurate representation of the evolving dynamics of the corresponding physical system. Moreover, the digital twin plays a pivotal role in various fields, such as meteorology, where it provides valuable insights for natural disaster prediction, maritime route navigation\cite{DT1}, agriculture monitoring\cite{DT2}, and city power management\cite{c2_3}.
So far, digital twins have primarily operated on digital computers by repeatedly sampling continuous signals and employing finite-depth machine learning models, such as ResNet\cite{he2016deep}, to model a single discrete time transition. However, they face substantial challenges related to data, model, and hardware architecture. These issues can be effectively addressed by our proposed digital twin using a memristive neural ODE solver.\par
As for data representation, conventional approaches typically discretize continuous signals and apply discrete numerical methods to simulate time-continuous real-world systems (upper data panel of \figref{fig:co}b). However, this method introduces truncation errors and information loss. In contrast, our solution features intrinsic continuous-time dynamics analogous to the physical space counterpart (bottom data panel of \figref{fig:co}b).\par
As for the model, recurrent ResNets  (upper model panel of \figref{fig:co}c) are frequently employed to parameterize the difference state evolution equation: \begin{math}h_{t+1}=h_t+f(h_t, \theta)\end{math},
where difference is parameterized by a ResNet $f$ with parameters $\theta$, such as a ResNet comprising multiple neural network blocks. Increasing the network depth by stacking more neural network blocks enhances ResNet's representation capability, which better models the complex dynamics of its physical world counterpart. However, this improvement significantly increases parameter population, challenging edge deployment. It also increases training complexity when traditional gradient descent methods are employed.\par
To augment representation capability and mitigate the training overhead of traditional AI-powered digital twins, we introduce neural ODEs. As shown in the bottom model panel of \figref{fig:co}c, neural ODE integrates multiple neural network blocks mirroring the ResNet, followed by a differential operator. Notably, a recurrent ResNet composed of an infinite number of identical ResNets can be conceptualized as the solution to a neural ODE that characterizes the continuous dynamics of hidden states within the digital twin, as described by\cite{neural_ode}:
\begin{equation}
    \centering
    \frac{{dh(t)}}{{dt}} = {f}(h(t),t, \theta)
    \label{eq:ode}
\end{equation}
The solution for the hidden state $h(t)$ at any given time can be computed using a black-box differential equation solver (see Methods). This approach outperforms finite depth models (e.g., recurrent ResNet) with the same parameter population. It also enables the training of the continuous-time digital twin using the low-cost adjoint method.\par
As for hardware architecture, traditional digital systems face significant challenges that profoundly impact their overall efficiency (upper architecture panel of \figref{fig:co}d). A key issue is that the frequent analogue-digital conversion leads to increased power consumption in the digital system\cite{hybrid}. Furthermore, as \gls{cmos} technology approaches its physical limits, Moore's Law\cite{co-moore1}, a driving force behind transistor development, encounters obstacles such as quantum effects, heat dissipation, and fabrication complexity\cite{co-moore2,co-moore3,co-heat}. Additionally, the well-known von Neumann bottleneck incurs overhead in data transfer between processing and memory units which exacerbates energy consumption and constrains computing speed\cite{co4}.\par
To overcome the aforementioned challenges, our system (bottom architecture panel of \figref{fig:co}d) leverages high-density memristor arrays. It features an in-memory computing architecture that integrates processing, memory and storage, thus overcoming the von Neumann bottleneck and enhancing computing parallelism as well as energy and area efficiency\cite{co_im_1,co_im_2}. Furthermore, our system is fully analogue, obviating the need for frequent analogue-digital conversions, thereby further reducing energy consumption and latency.

\section*{Memristive neural ordinary differential equation solver}
\begin{figure}[!t]
    \centering
    \includegraphics[width=0.9\linewidth]{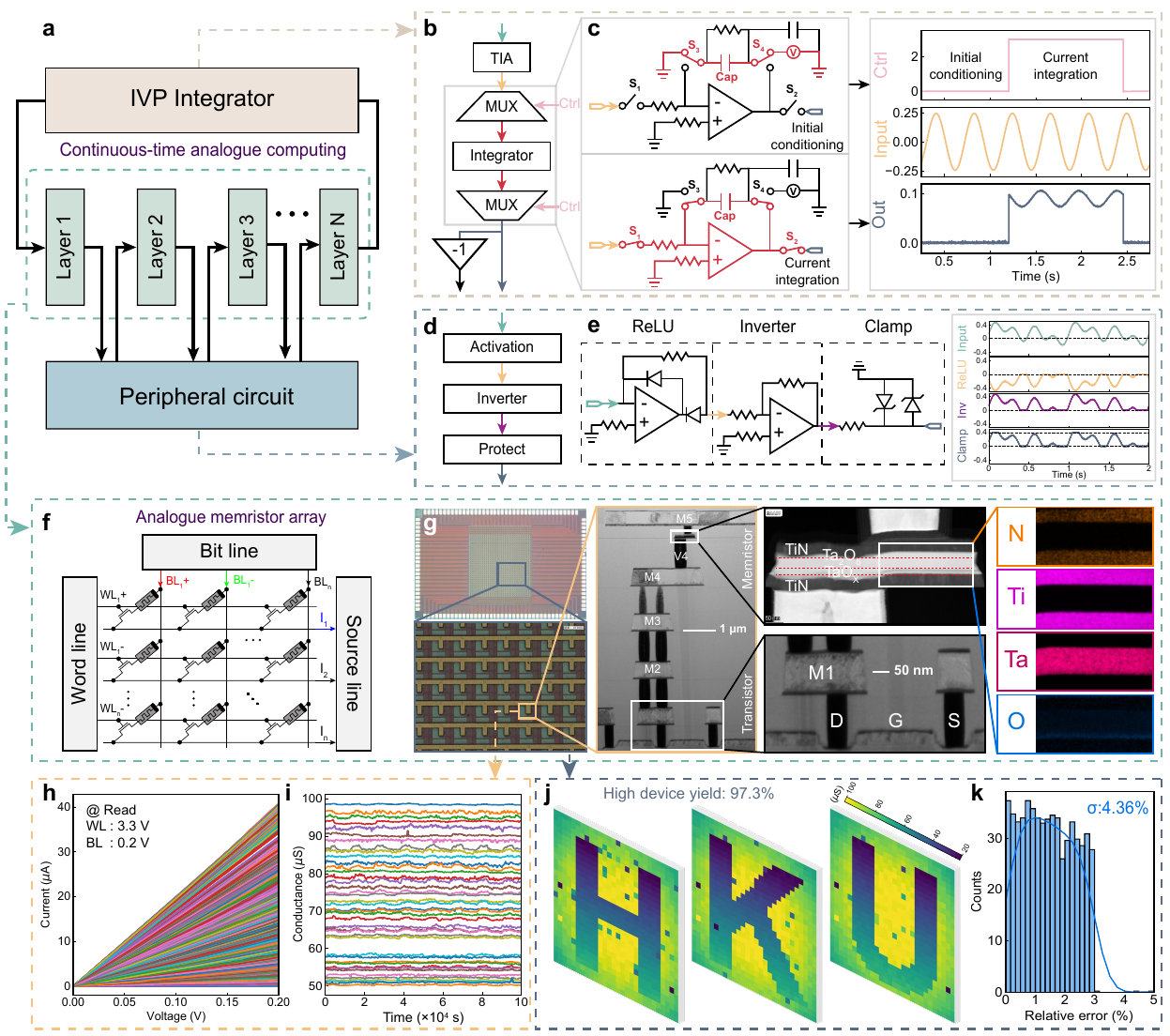}
    \caption{\textbf{Hardware implementation of memristive neural ordinary differential equation (ODE) solver and the characteristics of the analogue memristor array.} 
    \textbf{a,} System diagram, consisting of an IVP integrator, peripheral circuit, and analogue memristor arrays.
    \textbf{b,} The IVP integrator consists of a \gls{tia}, an integrator with analogue multiplexers, and a voltage inverter. It integrates output from the analogue memristive neural network and feeds back to the input of the analogue memristive neural network, equivalent to the differential operator of a neural ODE.
    \textbf{c,} The IVP integrator's dual modes: initial conditioning and current integration, with oscilloscope-acquired waveforms.
    \textbf{d,} The peripheral circuit provides analogue activation, current-to-voltage conversion, voltage inversion, and protection.
    \textbf{e,} The \gls{relu} and inverter in the periphery circuit convert currents to voltages for the next layer's inputs, protected by a clamp circuit. The right panel shows oscilloscope-acquired waveform.
    \textbf{f,} The analogue memristor array uses differential pairs for weight mapping to represent both positive and negative weights. 
    \textbf{g,} Optical photo of the \SI{180}{\nano\meter} 32$\times$32 analogue memristor array and magnified crossbar, as well as cross-sectional TEM of an analogue memristor, fabricated between the metal 4 and metal 5 layers. 
    \textbf{h,} Analogue programming of individual memristors shows more than 64 states.
    \textbf{i,} Retention exceeding $10^5$ seconds for different analogue conductance.
    \textbf{j,} Conductance map for the letters \textquotesingle H\textquotesingle, \textquotesingle K\textquotesingle, and \textquotesingle U\textquotesingle\ with a high device yield of 97.2\%.
    \textbf{k,} Corresponding relative programming error distribution, with a variance of 4.36\%. 
    }
    \label{fig:system}
\end{figure}
\figref{fig:system} illustrates the memristive neural ODE solver and the characteristics of the analogue memristor array.\par
In \figref{fig:system}a, the memristive neural ODE solver is a time-continuous and analogue computing system (see Methods for details and Supplementary Fig. 1 for the system photo). This system comprises three building blocks: the initial value problem (IVP) integrator, the peripheral circuit, and the analogue memristor array.\par
As illustrated in \figref{fig:system}b, the IVP integrator physically implements the integration operator, which is equivalent to the differential operator of a neural ODE (see Methods for the definition of the neural ODE). The IVP integrator operates in two modes: initial conditioning and current integration. The switching between these two modes is facilitated by analogue multiplexers.
To set the required initial condition for the IVP integrator, the integrating capacitor is pre-charged, as shown in \figref{fig:system}c. Opening switches $S_1$ and $S_2$ isolates the IVP integrator from external inputs, while closing switches $S_3$ and $S_4$ connects the capacitor to the power supply, charging it to a preset initial voltage to represent the initial condition of the neural ODE. Transitioning into the current integration mode involves toggling the states of all analogue multiplexers, which connects the IVP integrator to the analogue memristor arrays to solve the neural ODE as an IVP. \figref{fig:system}c also presents an example waveform captured by an oscilloscope, including the control signal (red), input signal (green) through the \gls{tia}, and integrator output (blue). Experimental integration results closely match those from the circuit simulation, with minimal error.\par
As shown in \figref{fig:system}d, the peripheral circuit performs analogue activation, including \gls{relu}\cite{relu}, current-to-voltage conversions, voltage inversion, and protection. The incorporation of a clamp circuit safeguards the system against over-voltage conditions. The right panel compares the experimental activation and clamped outputs with circuit simulation results.\par
\figref{fig:system}f illustrates the block diagram of the analogue memristor array. Each weight of the neural ODE is mapped to the conductance difference between two memristors configured as a differential pair. When the input voltage signals, denoted by the red and green lines, are applied to two adjacent columns with equal amplitude but opposite polarity, this enables the differential pair to encode both positive and negative weights.
The neural ODE is physically implemented using 32$\times$32 1T1R memristor arrays, as shown in \figref{fig:system}g. The memristors are integrated with \gls{cmos} on the \SI{180}{\nano\meter} standard logic platform, positioned between the metal 4 and metal 5 layers. The $\mathrm{TiN}/\mathrm{TaO}_x/\mathrm{Ta}_2\mathrm{O}_5/\mathrm{TiN}$ memristor is fabricated using the back-end-of-line process, as observed by a Transmission Electron Microscope (TEM) (see Supplementary Fig. 2 for more information about memristor characteristics). The \gls{eds} mapping results provide a detailed illustration of the elements distribution (Ti, N, Ta, O) across various layers. The unique dual-layer structure of $\mathrm{Ta}_2\mathrm{O}_5/\mathrm{TaO}_x$ is also revealed by the distribution of O (see Methods for further details on fabrication).\par
\figref{fig:system}h showcases the multi-level conductance of a single memristor, achieving high-precision 6-bit resolution, corresponding to over 64 distinct conductance states. \figref{fig:system}i demonstrates stable analogue conductance by repeatedly applying 0.2 V to selected memristors in the array for over 10,000 seconds.
\figref{fig:system}j shows three programmed letters on analogue memristor arrays, demonstrating a high yield of 97.3\% and reasonable programming accuracy at the array level (see Supplementary Fig. 3 and Fig. 4 for the programming scheme and numerical conductance of the letters).
\figref{fig:system}k presents the corresponding histogram depicting the distribution of programming errors, defined as the relative error between the target conductance and the post-programming conductance for responsive memristors (see Supplementary Fig. 5 for the analysis of patterns), showing a variance of 4.36\%.

\section*{Digital twin for the HP memristor}
\begin{figure}[!t]
    \centering
    \includegraphics[width=0.9\linewidth]{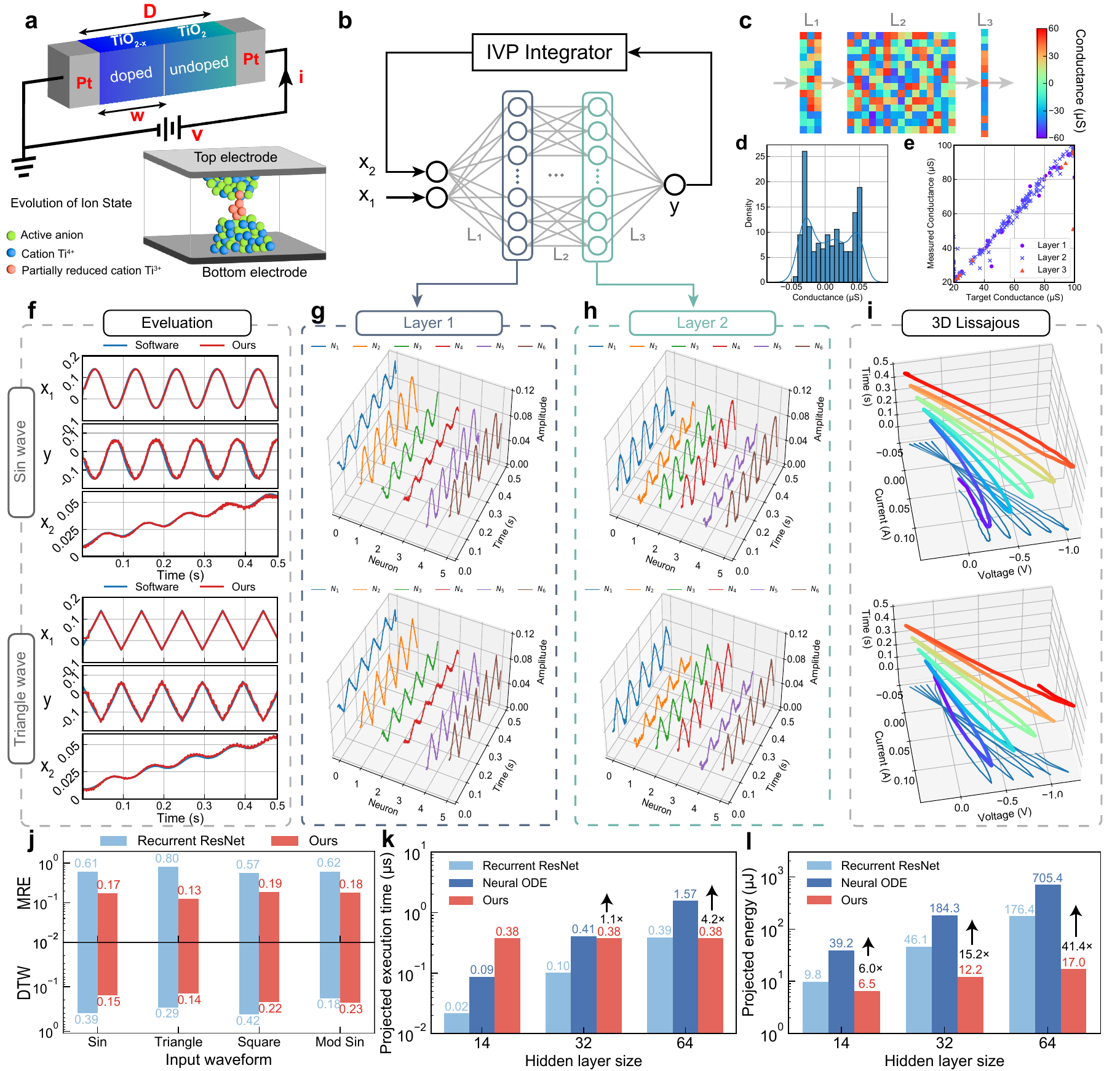}
    \caption{\textbf{Experimental digital twin of an HP memristor.}
    \textbf{a,} Schematic of the HP memristor, showing how the boundary between doped and undoped regions evolves under applied voltage.
    \textbf{b,} Framework of our system, with the neural ODE implemented on a three-layer analogue memristive neural network.
    \textbf{c,} Experimentally programmed differential conductance in the three analogue memristor arrays. 
    \textbf{d,} Histogram illustrating the distribution of programmed conductance.
    \textbf{e,} Statistical analysis showing less than 2.2\% error within the conductance range of \SI{20}{\micro\siemens} to \SI{100}{\micro\siemens}. 
    \textbf{f,} Comparison of experimental voltage waveforms from our system and software ground truth upon sinusoidal and triangular input stimulation.
    \textbf{g-h,} Experimental voltage waveforms of selected $L_1$ and $L_2$ hidden neurons during inference.
    \textbf{i,} Three-dimensional, time-dependent Lissajous plot depicting the nonlinear current-voltage (I-V) relationship, induced by sinusoidal and triangular input stimuation and their corresponding states.
    \textbf{j,} Comparison of modelling errors with recurrent ResNet-based digital twin across four stimulation waveforms.
    \textbf{k-l,} Comparison of speed and energy consumption between recurrent ResNet on digital hardware, neural ODE on digital hardware, and our system across different model sizes, demonstrating projected improvements of 4.2-folder and 41.4-folder, respectively, at a hidden size of 64.
 }
    \label{fig:hp}
\end{figure}
To validate our approach, we first construct a digital twin of a current-controlled memristor proposed by the Hewlett–Packard (HP) laboratories\cite{HP1}, which is characterized by a chaotic manifold governed by an ODE. \figref{fig:hp}a illustrates the dynamics of the HP memristor. Here, an external bias, $v(t)$, drives oxygen anions toward the electrode, thereby modifying the electrical conductivity through changes in the valence of metal cations. Assuming Ohmic electrical conduction, the resistance of the HP memristor follows\cite{HP2}:
\begin{equation}
\frac{v}{i} = R_{\text{ON}}\frac{w}{\text{D}} + R_{\text{OFF}}\left( 1 - \frac{w}{\text{D}} \right),
\label{eq_v_ron_roff}
\end{equation}
where $R_{\text{ON}}$ and $R_{\text{OFF}}$ are resistivities of the doped and undoped region, respectively. \text{D} is the memristor dimension between two metal terminals. The state variable of a HP memristor follows:
\begin{equation}
\frac{dw}{dt} = \mu_{\text{v}}\frac{R_{\text{ON}}}{D}i = f(w, v, t, \theta),
\label{eq_w_i}
\end{equation}
where $w$ is the state variable representing the boundary location between doped and undoped region with average ion mobility $\mu_v$. To model the HP memristor as a black box, we parameterize the right-hand side of the equation using a neural network $f$, forming a neural ODE.\par
\figref{fig:hp}b shows the schematic diagram of our system. The analogue input signals, $x_1$ and $x_2$, pass through the three-layer neural network physically implemented on three analogue memristor arrays (2$\times$14, 14$\times$14, 14$\times$1) and peripheral circuits. The output voltage $y$ from the last layer is integrated by the IVP integrator before serving as the input $x_2$ to the neural network. This closed-loop circuit represents the integral form of the neural ODE.
The weights of the neural networks are optimised offline before deployment on analogue memristor arrays (see Methods). \figref{fig:hp}c,d present the experimental conductance map of the differential pairs in three analogue memristor arrays, along with the distribution of conductance. \figref{fig:hp}e illustrates the low statistical programming errors of analogue memristor arrays, with an average relative error of 2.2$\%$.\par
To demonstrate that our digital twin behaves as a HP memristor described by Eq. (\ref{eq_w_i}), we tested it with four types of input stimulation (sine, triangular, rectangular, and modulated sine waveforms) and simulated the HP memristor's evolution. 
\figref{fig:hp}f compares results from our digital twin with software ground truth. Here, $x_1$, $y$ and $x_2$ are monitored using an oscilloscope. The experimental waveforms closely match the software-based ground truth, demonstrating the memristive digital twin's capacity to model, interpolate, and extrapolate complex HP memristor dynamics.
\figref{fig:hp}g,h show the experimental voltage outputs of six selected neurons in middle layers $L_1$ and $L_2$, respectively.
\figref{fig:hp}i presents a 3D time-dependent Lissajous plot of the I-V characteristic, depicting the non-linear relationship between the sinusoidal/trigonometric input $x_1$ and corresponding states $x_2$ over a continuous time span from 0 to 0.5 seconds.

\figref{fig:hp}j compares the modelling errors under different stimulation conditions between our system and a conventional digital twin using recurrent ResNet on digital hardware. Our approach achieves lower errors, with an \gls{mre} of 0.17 and a \gls{dtw} score of 0.15 (see Methods for the loss function definition). In contrast, recurrent ResNet on digital hardware exhibits higher errors, with a \gls{mre} of 0.61 and a \gls{dtw} score of 0.39 (see Supplementary Fig. 6 for the recurrent ResNet and neural ODE training process).
Next, we evaluate the speed and energy efficiency of our system through a scalability analysis against state-of-the-art Graphic Processing Units (GPUs)\cite{xie202116,lee2024neurosim}. As shown in \figref{fig:hp}k, a projected memristive neural ODE digital twin (same technology node and footprint) achieves 4.2$\times$ faster speeds than GPU-based neural ODE at a hidden layer size of 64. As the network scales up, the benefits of our system's time-continuous and analogue in-memory computing become more pronounced.
\figref{fig:hp}l presents the energy consumption comparison among three systems at varying hidden layer sizes. The light-blue and dark-blue bars indicate the estimated energy consumption of recurrent ResNet and neural ODE on a state-of-the-art GPU, respectively. Notably, our system (red bars) exhibits significantly lower energy consumption, approximately \SI{17.0}{\micro\joule} per forward pass. In comparison, at a  hidden layer size of 64, recurrent ResNet and neural ODE on a GPU consume around \SI{176.4}{\micro\joule} and \SI{705.4}{\micro\joule}, respectively. These results highlight a remarkable 10.4-fold and 41.4-fold improvement in energy efficiency compared to recurrent ResNet and neural ODE, respectively, both on digital hardware.

\section*{Digital twin for multivariate time series extrapolation using Lorenz96 dynamics}
\begin{figure}[!t]
    \centering
    \includegraphics[width=0.9\linewidth]{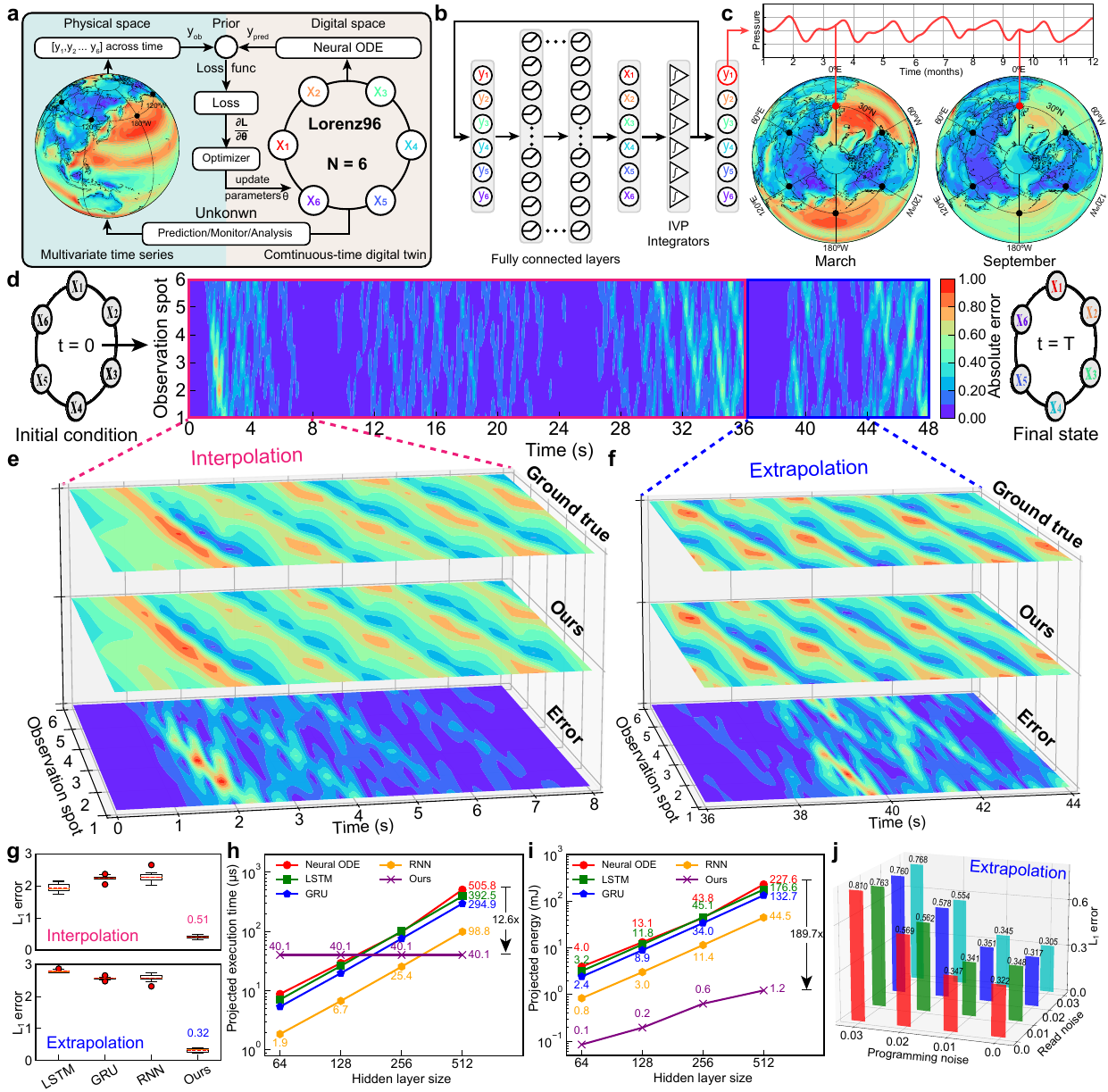}
\caption{\textbf{Multivariate time series extrapolation of Lorenz96 dynamics with our digital twin.}
    \textbf{a,} Illustration of training framework for our digital twin.
    \textbf{b,} The architecture of the digital twin, featuring fully connected layers and a group of six IVP integrators, facilitates the dynamic self-evolution of six-dimensional states ($y_1$ to $y_6$) without external stimulation.
    \textbf{c,} Temporal evolution of atmospheric pressure $y_1$ at $30^\circ$N, $0^\circ$E, modelled using Lorenz96 dynamics, depicting annual fluctuations in Earth's atmospheric pressure with detailed snapshots for March and September.
    \textbf{d,} Absolute error (\(L_1\)) between the output of our system and the ground truth, with the interpolation stage (0-36s) highlighted in red and the extrapolation stage (36-48s) marked in blue.
    \textbf{e-f,} Contour plots showcasing the selected interpolation (0-8s) and extrapolation (36-44s) phases, including the ground truth, output results from our system, and the \(L_1\) error.
    \textbf{g,} Comparison of \(L_1\) error in interpolation (red) and extrapolation (blue) for four different models: LSTM, GRU, and RNN on digital hardware and our system.
    \textbf{h-i,} Comparison of execution time and energy consumption for 5 systems (neural ODE, LSTM, GRU, and RNN on digital hardware as well as our system) executing an interpolation task from 0s to 36s, across increasing model sizes of 64, 128, 256, and 512.
    \textbf{j,} Noise analysis of our system, evaluating the impact of various combinations of read and programming noise during the inference. Our system is robust to read noise.
}
    \label{fig:Lorenz96}
\end{figure}
In addition to the HP memristor, we assess the performance of our digital twin in modelling the Lorenz96 dynamics, a simplified mathematical representation of atmospheric physics, which is widely used in atmospheric variability and climate prediction\cite{lo1,lo2}.
In \figref{fig:Lorenz96}a, we show the training process for the digital twin using physical space observations $y_{ob}$. The loss function quantifies the disparity between observed and predicted outcomes $y_{pred}$, enabling the digital twin to accurately mirror the state of the physical asset\cite{DT_city,DT_wea_1,time_pre}.\par
The Lorenz96 dynamics is described by a series of ordinary differential equations:
\begin{equation}
    \frac{dx_i}{dt} = \left( x_{i+1} - x_{i-2} \right) x_{i-1} - x_i + F, \quad i = 1, \ldots, n > 3,
\label{eq_Lorenz96}
\end{equation}
with a periodic boundary condition denoted as ${x_{i + n}} = {x_i}$, where $n \in \mathbb{N}$ and $F \in \mathbb{R}$ are the parameters. It represents atmospheric waves circulating along a latitude circle, with each segment ${x_i}$ representing a distinct meteorological variable such as pressure or precipitation, where each index $i$ indexes latitude segments.\par
\figref{fig:Lorenz96}b illustrates our system, which consists of a 4-layer fully connected neural network and six IVP integrators. Unlike the HP memristor modelling, the evolution of Lorenz96 dynamics is autonomous, without external stimulation.
\figref{fig:Lorenz96}c presents an example of the temporal evolution of atmospheric pressure ($y_1$) at $30^\circ$N and $0^\circ$E according to Lorenz96 dynamics. It also provides an overview of atmospheric pressure patterns for March and September.\par
The training and testing processes are similar to those used for the previous digital twin of the HP memristor. Despite its autonomous evolution without external input, this digital twin incorporates random noise as a regulariser during training to enhance its stability and robustness\cite{noise}. \figref{fig:Lorenz96}d and the expanded views \figref{fig:Lorenz96}e,f display the post-training error of our digital twin. The time axis is segmented into two phases: the interpolation phase (0-36 seconds) depicted by the red box, and the extrapolation phase (36-48 seconds) depicted by the blue box. The color indicates the magnitude of deviation between the digital twin's outcomes and the ground truth of the Lorenz96 dynamics, showing minimal discrepancy (see Supplementary Fig. 7 for more analysis). Additionally, the digital twin accurately predicts the non-linear dynamic behaviour across the seven largest Lyapunov times, while the influence of various hyperparameters is investigated (see Supplementary Note 1 and Supplementary Fig. 8 and 9 for detailed information on the training performance of neural ODE).\par
To benchmark the effectiveness of our digital twin, we conduct a series of performance evaluations against other multivariate time series models, including Long Short-Term Memory (LSTM), Gated Recurrent Unit (GRU), and Recurrent Neural Network (RNN), on the state-of-art GPUs. These models are compared on the basis of error, computational speed, and energy efficiency, while maintaining consistent parameter settings and training conditions (see Supplementary Table 1 for detailed information on speed and energy consumption).\par
In \figref{fig:Lorenz96}g, the errors over ten trials are shown, highlighting the mean performance and variation between different models on the interpolation (red) and extrapolation (blue) tasks. The average \(L_1\) error of our neural ODE digital twin for the interpolation task is approximately 0.512, while for the extrapolation task, it is approximately 0.321. In contrast, models such as LSTM, GRU, and RNN exhibit significantly larger errors (see Supplementary Fig. 10 for hidden layer size impact).\par
\figref{fig:Lorenz96}h presents the execution time through a scalability analysis for a single inference sample across five models with increasing hidden layer size. The estimated execution time for the neural ODE, LSTM, GRU, and RNN models on state-of-the-art GPU architecture is \SI{505.8}{\micro\second}, \SI{392.5}{\micro\second}, \SI{294.9}{\micro\second}, and \SI{98.8}{\micro\second}, respectively, with 512 hidden neurons. In comparison, the projected execution time of a neural ODE on analogue memristors is \SI{40.1}{\micro\second}, which represents enhancements of approximately 12.6-fold, 9.8-fold, 7.4-fold, and 2.5-fold over the neural ODE, LSTM, GRU, and RNN models on digital hardware, respectively (see Supplementary Note 2 for the projected speed estimation).
\figref{fig:Lorenz96}i presents the energy consumption of each model across four different hidden layer sizes, with 64, 128, 256, and 512 hidden neurons. The memristive neural ODE demonstrates projected energy efficiency improvements, surpassing the neural ODE, LSTM, GRU, and RNN models on digital hardware (512 hidden neurons) by factors of 189.7, 147.2, 100.6, and 37.1, respectively (see Supplementary Note 2 for the projected energy estimation).\par 
Analogue circuits often encounter noise issues, which can lead to reduced accuracy and unexpected signal fluctuations. \figref{fig:Lorenz96}j examines the effect of read and programming noise on the performance of our digital twin in extrapolation tasks. Our findings, each averaged over ten repetitions, indicate that read noise can even reduce extrapolation errors compared to a noise-free environment. For instance, with read noise at 2\% and no programming noise, the model achieves a lower \(L_1\) error of 0.317, in contrast to 0.322 without read noise. This highlights the resilience of our digital twin to read noise.\par

\section*{Conclusions}


In this work, we present a continuous-time and in-memory neural ODE solver for infinite-depth AI-powered digital twins. This approach effectively addresses the data, model and architecture limitations of conventional digital twins. Through validation across two tasks-modelling HP memristor and extrapolating Lorenz96 dynamics, our methodology shows significant improvements.
The experimental digital twin for HP memristor demonstrate a substantial enhancement over the conventional digital twins, with our system achieving a 4.2-fold projected increase in speed and a 41.4-fold projected reduction in energy consumption. Moreover, in extrapolating Lorenz96 dynamics, our system achieves speed improvements of 12.6-fold, accompanied by energy savings of 189.7-fold when benchmarked against neural ODEs on digital hardware.
Our system holds great promise for advancing efficient and accelerated digital twin technology, benefiting Industry 4.0 initiatives. 

\section*{Method}

\subsection*{Fabrication of analogue memristor chips}

In this work, a 1Kb analogue memristor chip, comprising a 32$\times$32 1T1R crossbar array, is fabricated using the \SI{180}{\nano\meter} technology node. Following the deposition of the Via4 layer during the backend-of-line process, memristors are stacked on the drain side of the transistors. Initially, a \SI{40}{\nano\meter} TiN layer, serving as the bottom electrode, is formed through Physical Vapor Deposition (PVD). Subsequently, a dual TaO-based dielectric layer with different oxygen compositions is designed and deposited by Atomic Layer Deposition (ALD) to enhance the analogue properties of memristors. The dielectric layer included a \SI{50}{\nano\meter} lower-O-concentration $\mathrm{Ta}_2\mathrm{O}_5$ layer and a \SI{10}{\nano\meter} higher-O-concentration $\mathrm{TaO}_x$ layer. Then, a \SI{40}{\nano\meter} TiN layer, fabricated by PVD, is used as the top metal. Finally, the top Metal 5 process is completed to finish the fabrication. The 1T1R crossbar structure is implemented by connecting the gate and source terminals of the transistors as the column-wise \gls{wl} and \gls{sl}, respectively, and the TE terminals of the individual memristor as the row-wise \gls{bl}. The device underwent a post-annealing treatment under vacuum conditions at a temperature of \SI{400}{\celsius} for a duration of 30 minutes. This treatment significantly improved the performance of the chip, resulting in devices that exhibit high endurance and reliability.

\subsection*{The fully analogue computing system}
The fully analogue computing system comprises three \SI{180}{\nano\meter} analogue memristor arrays, each equipped with selection transistors. Additional components include switch matrix boards and peripheral circuits, which facilitate comprehensive system integration and functionality. The system is controlled via a Personal Computer (PC) and an ARM MCU STM32F407ZGT6 (see Supplementary Fig. 1). The analogue memristor array operates in two modes: programming mode, used for weight mapping of the analogue memristor array, and multiplication mode, which supports the vector-matrix multiplication.
\subsubsection*{Programming mode}
In programming mode, the target conductance within the analogue memristor array is meticulously adjusted to align with the software-defined weight values. The selection of the specific memristor for programming is controlled by a switch matrix that activates analogue multiplexers (TMUX) for the \gls{wl}, \gls{bl}, and \gls{sl}. This configuration directs the routing for the targeted memristor selection transistors to a low-resistance state, ensuring precise control while keeping other memristors in a high-resistance state to prevent interference. Isolating the selected memristor allows for direct interfacing with the \gls{b1500}, facilitating accurate and efficient programming. The adjustment process is executed via a Python script on a PC, which interfaces with the \gls{b1500} to finely tune the device parameters, thereby achieving the desired conductance levels that accurately reflect the neural network weights in the analogue memristor (see Supplementary Fig. 3 for details of the programming scheme).\par
\subsubsection*{Multiplication mode}
In the multiplication mode, a PC commands the switch matrix boards, controlled by MCUs, to establish connections with the peripheral circuitry. It uses a waveform generator and oscilloscope to administer and record voltages, respectively. To facilitate vector-matrix multiplication, analogue voltages generated by the waveform generator are routed to the analogue memristor array's bit lines through a dual-channel analogue multiplexer (TMUX1134; Texas Instruments, TI). The resulting currents on the source lines are fed into trans-impedance amplifiers (\gls{tia}s) (OPA4990; Texas Instruments, TI), which convert the currents into voltages. Voltage activation is achieved through a \gls{relu} module utilizing a dual-diode setup (with diode 1N4148) within the \gls{tia}. These voltages are reversed and integrated by inverting amplifier and integrator (OPA4990; Texas Instruments, TI). The voltage outputs are then fed back as inputs to the analogue neural network block, creating a closed-loop system that emulates the neural ordinary differential equation. The voltage at the integrator is captured by an analogue-to-digital converter (ADC) (ADS781; Texas Instruments, TI), which transmits the data back to the PC for collection and analysis.
\subsection*{Mean relative error}
By averaging the relative errors across all data points, we obtain a comprehensive measure that reflects the overall predictive performance. For two time series, the predicted series $X = \{ {x_1},{x_2}, \ldots ,{x_n}\} $ and the ground truth series $Y = \{ {y_1},{y_2}, \ldots ,{y_n}\}$, the mean relative error (MRE) for the entire series can be calculated as follows:
\begin{equation}
    MRE(X,Y) = \frac{1}{{n}}\sum\limits_{i = 1}^n {\left| {\frac{{{x_i} - {y_i}}}{{{y_i}}}} \right|} 
    \label{RE}
\end{equation}
MRE provides a normalized assessment of the error, accounting for the scale of the data. This is particularly important when analysing time series data that exhibit significant fluctuations in magnitude. By incorporating the scale of the data, the MRE offers a more accurate evaluation of the error, enabling precise comparisons and analysis.
\subsection*{Dynamical time wrapping}
DTW is a versatile algorithm widely employed to measure the similarity between two temporal sequences that may differ in speed or length\cite{softdtw, dtw}. Let's consider two time series, $X = \{ {x_1},{x_2}, \ldots ,{x_n}\} $ and $Y = \{ {y_1},{y_2}, \ldots ,{y_m}\}$.To apply \gls{dtw}, we construct an $n \times m$ matrix where each element $d_{i,j}$ represents the distance between the $i^{th}$ element of X and the $j^{th}$ element of Y, typically using the Euclidean distance:
\begin{equation}
\centering
    {d_{i,j}} = \left| {{x_i} - {y_j}} \right|
    \label{dtw_dis}
\end{equation}
Then, we compute the cumulative distance $D_{i,j}$ for each element of the matrix using dynamic programming, following the recursive relation:
\begin{equation}
\centering
    {D_{i,j}} = {d_{i,j}} + \min ({D_{i - 1,j}},{D_{i,j - 1}},{D_{i - 1,j - 1}})
    \label{dtw_ite}
\end{equation}
The objective of \gls{dtw} is to find a path from $(1,1)$ to $(n,m)$ that minimizes $D_{i,j}$, representing the total match cost between the two time series. To achieve this, we initialise $D_{0,0}$ to 0 and set $D_{0,j}$ and $D_{i,0}$ to infinity (or a sufficiently large number) to ensure that the matching path always starts at $(1,1)$. The recursive relation allows us to calculate $D_{i,j}$ based on the minimum of the three neighbouring elements: $D_{i-1,j-1}$, $D_{i-1,j}$, and $D_{i,j-1}$. This approach enables the elastic transformation of the time axes, allowing for optimal matching between the sequences, even when they vary in speed or length.

\subsection*{Comparison of ResNet and neural ODE architectures}
The residual neural network (ResNet), composed of an infinite number of identical neural network blocks, is defined as follows\cite{he2016deep}:
\begin{equation}
\centering
    h_{t+1}=h_t+f(h_t, \theta),
    \label{eq:resnet}
\end{equation}
where the gradient $f$ is parameterized by a neural network with parameters $\theta$. Skip connections enhance the ResNet's ability to learn residual features, thereby improving training and performance. These iterative updates can be interpreted as an Euler discretization of a continuous transformation.\par

On the contrary, neural ODEs describe continuous evolution using an ordinary differential equation specified by a neural network $f$, formulated as follows\cite{neural_ode}:
\begin{equation}
    \centering
    {h_{{t_1}}},{h_{{t_2}}} \ldots ,{h_{{t_N}}} = \text{ODESolve}({h_{{t_0}}},f,\theta ,{t_0}, \ldots ,{t_N})
    \label{eq:odesolve}
\end{equation}
Given observation times ${t_0}, \ldots, {t_N}$ and an initial state ${h_{t_0}}$, an ODE solver computes ${h_{t_1}}, {h_{t_2}}, \ldots, {h_{t_N}}$, representing the latent state at each observation.

\subsection*{Training method of continuous-time digital twin}
To train the neural ODE model $f$, we utilize the adjoint state method\cite{neural_ode}. This method computes the gradient of the loss function $L$ with respect to the hidden state at each time stamp, known as the adjoint $a(t) = \frac{{\partial L}}{{\partial {h_t}}}$. By defining the state vector carefully, we can compute the necessary integrals to solve for $a(t)$ and $\frac{{\partial L}}{{\partial \theta}}$ in a single call to the ODE solver.
\subsubsection*{HP memristor modelling}
We create a training set called $y_{\text{true}}$, comprising 500 data points sampled from Eq. \eqref{eq_v_ron_roff} with a time interval of $\Delta t = 1 \times 10^{-3} \text{ s}$. Our objective is to minimize the \(L_1\) error between these points and the corresponding trajectories predicted by our digital twin, denoted as $y_{\text{pred}}$, throughout all time steps. Regarding the neural network architecture, we employ the \gls{relu} activation function for all layers except the output layer.
\subsubsection*{Multivariate time series extrapolation}\label{methods_subsec3_atmospheric pressure}
To train our digital twin for extrapolating the MTS dataset, we generate the Lorenz96 dynamics with a sequence length of 2400. The first 1800 data points are used for training (interpolation task), while the remaining points are used for testing (extrapolation task). We train a digital twin with dimension $d$=6 using a three-layer neural network with 64 neurons in each hidden layer. Firstly, we set the initial conditions as \([y_{1}, y_{2}, \ldots, y_{6}] = [-1.2061, 0.0617, 1.1632, -1.5008, -1.5944, -0.0187]\). The model is optimized using the 0.02 millisecond trajectory of the Lorenz96 dynamics. We employ the \gls{dtw} as the loss function to quantify the dissimilarity between the predicted trajectories and the true trajectories ($y_\text{true}$). Our objective is to minimize the loss by calculating the gradients of the loss with respect to the parameters $\theta$ using the adjoint method. To optimize the parameters $\theta$ and minimize the loss, we employ the Adaptive Moment Estimation (Adam) optimization algorithm, which adaptively adjusts the learning rate. The hidden weights are updated using the loss gradients obtained from the \gls{dtw}, with a fourth-order Runge-Kutta solver (RK4) method serving as the ODESolve. The training process continues until the \gls{dtw} approaches zero or reaches the maximum number of epochs, indicating a close alignment between the predicted and true trajectories.\par
\subsubsection*{Lyapunov time}
We assess the extrapolation capability of our digital twin by using the Lyapunov time\cite{c1_3}, a critical metric for assessing predictability in chaotic systems. The Lyapunov time acts as a benchmark, indicating the period over which our forecasts are expected to retain a significant degree of accuracy. This measure of unreliability arises from the exponential amplification of initial errors, which is encapsulated by the system's maximal Lyapunov exponent (MLE). This exponent is defined as follows:
\begin{equation}
    \lambda ({x_0}) = \mathop {\lim }\limits_{n \to \infty } \frac{1}{n}\sum\limits_{i = 0}^{n - 1} {\ln } \left| {f'({x_i})} \right|,
    \label{lyapunov_time}
\end{equation}
where $f$ represents the function governing the chaotic dynamics and $n$ denotes the data points. Consequently, the lyapunov time is inversely proportional to the MLE.


\section*{Acknowledgements}
This research is supported by the National Key R\&D Program of China (Grant No. 2022YFB3608300), National Natural Science Foundation of China (Grant Nos. 62122004, 62374181), Hong Kong Research Grant Council (Grant Nos. 27206321, 17205922, 17212923). This research is also partially supported by ACCESS – AI Chip Center for Emerging Smart Systems, sponsored by Innovation and Technology Fund (ITF), Hong Kong SAR.


\section*{Competing interests}
The authors declare no competing interests.

\section*{Data availability}
The HP memristor and Lorenz96 dynamics data are included with this paper and can also be accessed at the provided GitHub repository: \url{https://github.com/chenhg99/NODEs_digital_twin}. For access to any other measured data mentioned in the paper, please feel free to request it.

\section*{Code availability}
All simulations involving neural ODEs were conducted using the PyTorch framework. The source code used in this study is openly accessible on the GitHub repository: \url{https://github.com/chenhg99/NODEs_digital_twin}.

\bibliography{bibliography}
\end{document}